\documentclass[]{article}
\usepackage{graphics}
\usepackage{graphicx}
\usepackage{amsfonts}
\usepackage{amssymb}
\usepackage{cite}
\usepackage{xcolor}
\usepackage{dsfont}
\def\beqn{\begin{eqnarray}}
\def\eeqn{\end{eqnarray}}
\def\aap{a^\dagger}
\def\aam{a}
\def\aapt{\tilde a^\dagger}
\def\aamt{\tilde a}
\def\hw{\hbar \omega}
\def\hw4{ \frac {\hbar \omega}{4}}

\def\uni{{\bf i}}
\def\re{{\rm {e}}}

\def\al{\alpha}

\def\Om{\Omega}
\def\om{\omega}
\def\b0{b_0}

\def\hu{\hat{u}}

\def\hv{\hat{v}}

\title{Psedo-hermitian Hamiltonians at finite temperature}
\author{
Romina Ram\'\i rez$^{1}$ and Marta Reboiro$^{2}$ $^*$ \\
\small $^{1}$ IAM, CONICET-CeMaLP, University of La Plata, Argentina. \\
\small $^{2}$ IFLP, CONICET-Dept. of Phys., University of La Plata, Argentina. \\
\small $^{*}$Corresponding author: M.R.; \tt{reboiro@fisica.unlp.edu.ar}
}

\date{}

\begin{document}
\maketitle
\begin{abstract}
The Double Green Function Formalism has been extensively used in dealing with the thermodynamics of quantum systems which evolved in time under the action of a given self-adjoint Hamiltonian. In this work, we extend the formalism to include pseudo-hermitian Hamiltonians. We apply the formalism to study the PT-symmetry Swanson Hamiltonian at finite temperature, both in the PT-unbroken and in the PT-broken symmetry phase. We analyse the behaviour of the system, which is initially at equilibrium at a given temperature, when it is perturbed by a periodic interaction.
\end{abstract}

%
\vspace{2pc}

%


\section{Introduction.}

The Field of Quantum Thermodynamics has increasingly gained attention in the last years \cite{book1,book2}. There have been both theoretical \cite{jarzynski0,jarzynski1,jarzynski2,jarzynski3,jarzynskipt,qt1,entropy2,entropy,theory1} as well as experimental advances \cite{schmidt0,schmidt1,schmidt2,zanin}. Thermodynamics is based on equilibrium states \cite{feynman},  the way a quantum system reaches an equilibrium state depends on the quantum dynamics of the system and its environment \cite{book1}. Among other works, the quantum thermodynamics of open systems has been discussed in \cite{out0,open1,open2}. Some novel articles analyse non-equilibrium thermodynamics problems by modelling the dynamics of the quantum system by non-hermitian Hamiltonians \cite{rotter,out1,out2,out3,out4,gasesnh,dgf-new}. Non-hermitian Hamiltonians have been proved to be useful in different physical contexts, i. e. gain and loss physics \cite{gainloss}, optics and photonics \cite{optics}, condensed matter \cite{condensed}. They arise as an approach that consists in modelling the dynamics of the system, in interaction with its environment, by an effective non-hermitian Hamiltonian, i.e. Feshbach Projection Method \cite{feshbach,feshbach1}. As an example, the Authors of \cite{out4} have analysed the role of fluctuations in a generalised  tight-binding lattice model. In the same direction, transport phenomena in open non-hermitian nano-structures has been studied in \cite{out3}. 

A well known formalism that allows us to deal with processes out of equilibrium is Double-time temperature-dependent Green Function (DGF) method \cite{ter-haar,green-1,green-2,green-3,arrachea,dgf-new}. This formalism has been extensively used in the study magnetic and conduction problems \cite{green-1,green-2,green-3}. More recently, the DGF has been used to solve transport problems \cite{ arrachea,dgf-new}. Among the available literature, the Author of \cite{arrachea} has presented a general treatment, based on non-equilibrium Green functions, to study transport phenomena in systems described by tight-binding Hamiltonians coupled to reservoirs. Moreover, the Authors of  \cite{dgf-new} have proposed the study of a condensate system with a boundary-dependent dynamical instability, they carry out the analysis of its dynamical properties in the framework of the Double Green Function formalism.

Amid other pseudo-hermitian systems\cite{ali1}, the Swanson model \cite{swanson} and its generalisations have been investigated in different contexts \cite{sw0,sw1,sw2,sw3,sw4,sw5,sw6,sw7,sw8,sw9,sw10,sw11,sw12,sw13,sw14,nos1,nos2}. The Swanson Hamiltonian obeys Parity-Time Reversal ({\cal{PT}}) symmetry \cite{bender1,bender2,bender3,bender4,bender5,bender6}  and it combines both physical and mathematical advantages. From the mathematical point of view, the Swanson model is described by a quadratic Hamiltonian. From the physical point of view, it can be mapped, through appropriate similar transformations, to a harmonic oscillator (HO) or to an inverted harmonic oscillator (IHO). Thus, depending on the parameters of the model, it can be used to model bound systems (HO) in the PT-symmetry phase, or systems with instabilities (IHO) in the PT-broken symmetry phase. Particularly, the study of the PT-broken symmetry phase of the Swanson model becomes relevant due to the renewed interest in the physics of IHO. As reported in \cite{iho0} the IHO Hamiltonian can be studied from different points of view, i.e. a dilatation generator, a squeeze generator, a Lorentz boost generator, or a scattering potential. The Authors of \cite{iho0} demonstrate that the physics of the IHO underlies phenomena so different as the Hawking–Unruh effect and the scattering in the lowest Landau level of quantum Hall systems. 

The aim of the present work is to extend the formalism of the DGF to include pseudo-hermitian Hamiltonians, so to describe the thermodynamic properties of a system that evolves in time at finite temperature. 

The work is organised as follows. In Section \ref{formal}, we review the time evolution formalism for pseudo-hermitian Hamiltonians. In \ref{sdgf} we formalise the extension of the DGF to include pseudo-hermitian Hamiltonians. In \ref{results}, we  apply the formalism to study the PT-symmetry Swanson Hamiltonian at finite temperature, both in the PT and in the PT-broken symmetry phase. We present examples for the PT-unbroken phase in \ref{ej1} and \ref{ej2}. In \ref{ej3} we discuss a problem in the PT-broken symmetry phase, in connection with the physics of the IHO. 
Conclusions are drawn in Section \ref{conclusions}.

\section{Formalism.}
\label{formal}

We shall consider a pseudo-hermitian Hamiltonian,  ${\mathrm H}$, and its adjoint, ${\mathrm H}_c$. 
In general,  ${\mathrm H}$ and ${\mathrm H}_c$ have not eigenfunctions in ${\cal L}^{2}({\mathbb R})$. To overcome this difficulty we shall introduce a Gel'fand Triplet.

Let us briefly review the construction of the Gel'fand triplet \cite{gelfand,bhom}. We need a Hausdorff vector space with a convex topology and a scalar product, $(\Psi,\tau)$ to describe a quantum system. The completion of $(\Psi,\tau)$ is given by the Hilbert space with the topology $\tau_H$, $({\mathcal{H}}, \tau_H)$. We have to introduce a  completion of $(\Psi,\tau)$, with a finer topology $\tau_\Phi$, $(\Phi,\tau_\Phi)$, so that $\Phi \subset {\mathcal H} \Rightarrow {\mathcal H}^* \subset \Phi^*$. Here, $\Phi^*$ is the dual space of $\Phi$. After that, we introduce the anti-dual space of $\Phi$, $\Phi^\times$. In this form, we obtain the Gel'fand triplet

\beqn
\Phi \subset {\cal H} \subset \Phi^\times.
\label{gelfand3}
\eeqn
We shall denote the extensions operators ${\mathrm H}$ and ${\mathrm H}_c$ on $\Phi^\times$ as 
${\mathrm H}^\times$ and ${{\mathrm H}_c}^\times$, respectively. 

In what follows, we shall work with generalised functions of $\Phi^\times$ when necessary. Otherwise, we shall work with functions of the usual Hilbert space, $\cal H$.

We shall denote as ${\widetilde \phi}_\nu(x,t)$  the wave function associated to ${\mathrm H}^\times$,  and ${\overline \phi}_\nu(x,t)$ to the wave function of ${\mathrm H}_c^\times$, respectively. Namely:

\beqn 
{\bf i} \hbar ~\frac{\partial {\widetilde \phi}_\nu}{\partial t} &=& {\mathrm H}^\times {\widetilde \phi}_\nu, \nonumber \\
{\bf i} \hbar ~\frac{\partial {\overline \phi}_\nu}{\partial t} &=& {\mathrm H_c}^\times {\overline \phi}_\nu.
\label{eigen}
\eeqn 
As pointed out in \cite{sw0,fring2,fring3,fring4}, when the Hamiltonian varies with time, ${\mathrm H}^\times={\mathrm H}^\times(t)$, it can be related to a self-adjoint operator, ${\mathfrak h}^\times$, through

\beqn  
 {\mathfrak h}^\times= \Upsilon {\mathrm H}^\times \Upsilon^{-1}- {\bf i} \hbar \Upsilon \frac{\partial \Upsilon^{-1}}{\partial t}.
 \label{ht}
\eeqn 
Thus

\beqn 
{\mathrm H_c}^\times~S= S~{\mathrm H}^\times+ {\bf i} \hbar \frac{\partial S}{\partial t},
\label{simi}
\eeqn 
being $S$ the Dyson operator $S= \Upsilon^\dagger \Upsilon$. Notice that for $\Upsilon$ independent of the time, $t$, $S$ serves as a similar transformation between ${\mathrm H}^\times(t)$ and ${\mathrm H}_c^\times(t)$.

As proved in \cite{fring2}, the wave functions of ${\mathfrak h}^\times$, $\phi(x,t)$, are related to those of ${\mathrm H}^\times(t)$ and ${\mathrm H}_c^\times(t)$ by

\beqn 
{\widetilde \phi}_\nu(x,t) & = & \Upsilon^{-1} \phi_\nu(x,t), \nonumber \\
{\overline \phi}_\nu(x,t) & = & \Upsilon^\dagger \phi_\nu(x,t),
\eeqn 
respectively. 

In Schr\"odinger's representation, $\phi(x,t)$ evolves in time as $\phi(x,t)= \hat U(t,t_0) \phi(x,t_0)$. 
The time evolution operator $\hat U(t,t_0)$ obeys the equation  

\beqn 
{\bf i} \hbar \frac {\partial {\hat U}(t,t_0)} {\partial t} = {\mathfrak h}^\times {\hat U}(t,t_0),
\eeqn 
with the condition that $\hat U(t_0,t_0)=\hat I$, being $\hat I$ the identity operator.
Consequently

\beqn 
{\widetilde \phi}_\nu(x,t) & = & 
\Upsilon^{-1}(t) \hat U(t,t_0) \Upsilon(t_0)~{\widetilde \phi}_\nu(x,t_0)=  \hat {\widetilde U}(t,t_0) ~{\widetilde \phi}_\nu(x,t_0), \nonumber \\
{\overline \phi_\nu}(x,t) & = & 
\Upsilon^{\dagger}(t) \hat U(t,t_0) {\Upsilon^\dagger}^{-1}(t_0)~{\overline \phi}_\nu(x,t_0)=  \hat {\overline U}(t,t_0) ~{\overline \phi}_\nu(x,t_0).
\eeqn
It is straightforward to prove that $ \hat {\widetilde U}(t,t_0)$ and $\hat {\overline U}(t,t_0)$ obey the following equations

\beqn 
{\bf i} \hbar \frac {\partial {\hat {\widetilde U}}(t,t_0)} {\partial t} & = & {\mathrm H}^\times {\hat {\widetilde U}}(t,t_0), \nonumber \\
{\bf i} \hbar \frac {\partial {\hat {\overline U}}(t,t_0)} {\partial t} & = & {\mathrm H}_c^\times {\hat {\overline U}}(t,t_0),
\eeqn  
with $\hat {\widetilde U}(t_0,t_0)=\hat {\overline U}(t_0,t_0)=\hat I$.

From now on, and to obtain the main values of a given observable as a function of time, we shall adopt the Heisenberg representation. 

The density operator $\hat \rho(t)$, in the Heisenberg's representation, reads

\beqn 
\hat \rho(t)= {\hat U}(t,t_0) \hat \rho_0 {\hat U}(t,t_0)^\dagger,
\eeqn
and it can be obtained from the equation of motion

\beqn 
{\bf i} \hbar \frac{\rm {d} \hat \rho(t)}{\rm d t} = [ {\mathfrak h}^\times, \hat \rho(t)],
\eeqn 
with $\hat \rho_0= \hat \rho (t_0)$.
In this scheme, the mean value of an self-adjoint operator $\hat o$, as a function of time, is given by

\beqn 
\langle \hat o \rangle = {\rm {Tr}}_{(\mathfrak h)} ( \hat \rho(t) \hat o).
\eeqn 

To proceed we shall introduce the operator $\hat {\widetilde \rho}(t)$:

\beqn 
\hat {\widetilde \rho}(t) = {\hat {\widetilde U}}(t,t_0)  \hat {\widetilde \rho}(t_0) {\hat {\overline U}}^\dagger(t,t_0),
\eeqn
with  $\hat {\widetilde \rho}(t_0)= \Upsilon^{-1}(t_0) \hat \rho_0 \Upsilon (t_0)$. As we have proved in Appendix A the equation for $\hat {\widetilde \rho}$ is 

\beqn 
{\bf i} \hbar \frac{\rm {d} \hat {\widetilde \rho}(t)}{\rm d t} = [\hat {\mathrm H}^\times, \hat {\widetilde \rho}(t)],
\eeqn 
and 
the mean value of an operator $\hat O$, related to the self-adjoint operator $\hat o$ by $\hat O = \Upsilon^{-1}(t) \hat o \Upsilon(t)$ is 

\beqn 
{\overline{\hat O(t)}} = {\rm {Tr}}_{(\mathrm H^\times)} ( \hat {\widetilde \rho}(t) \hat O)={\rm {Tr}}_{(\mathfrak h)} ( \hat \rho(t) \hat o)= {\overline{\hat o(t)}}.
\label{ot}
\eeqn 

The object of the present work is to calculate the mean value of a physical observable when $\hat \rho_0$ is the equilibrium density operator at temperature $T$, that is

\beqn 
\hat \rho_0= \frac{\re^{-\beta~\mathfrak h_0}}{{\rm {Tr}}(\re^{-\beta~\mathfrak h_0})},
\eeqn 
with $\beta=(k_B T)^{-1}$ and $k_B$ the Boltzmann constant. In the present approach, we shall make use of the Double-Green Function \cite{green-1,green-2,green-3,ter-haar}, which we shall  
briefly review. 

\subsection{The formalism of Double Green Function.}\label{sdgf}

Let us begin by introducing the retarded, $G_r(t,t')$, and the advance,$G_a(t,t')$, Green Functions \cite{ter-haar}. Namely

\beqn 
G^r(t,t')=\langle \langle A(t); B(t') \rangle \rangle^r & =  
-& {\bf i}  ~\Theta(t-t') \langle [A(t), B(t')]\rangle,
\nonumber \\
G^a(t,t')=\langle \langle A(t); B(t') \rangle \rangle^a & = 
 & {\bf i} ~ \Theta(t'-t) \langle [A(t), B(t')]\rangle.
\label{green1}
\eeqn 
In Eq. (\ref{green1}), $\Theta(t)$ is the usual Heaviside function, and $\langle \hat{O}(t) \rangle$ is the thermal average of $\hat{O}(t)$ in the Canonical Ensemble

\beqn
\langle \hat{O}(t) \rangle & = & \frac 1Z 
\sum_n~\langle {\widetilde \phi}_n | \hat{O}(t) \re^{-\beta {\mathrm H}} |{\widetilde \phi}_n\rangle_S.
\label{aver}
\eeqn 
The partition function, $Z$, is written as

\beqn 
Z & =& {\sum_n~\langle {\widetilde \phi}_n | \re^{-\beta {\mathrm H}} |{\widetilde \phi}_n\rangle_S}.
\label{Z}
\eeqn
The inner products appearing in Eqs. (\ref{aver}) and (\ref{Z}) are taking respect to the metric operator $S$.

From Eqs. (\ref{green1}) and (\ref{aver}) it is clear that at $T=0$, we recover the usual Green Function.

After a few steps, the equation of motion, for both the retarded and the advance Green Function, is given by

\beqn 
{\bf i} \hbar \frac {{\rm d} G}{{\rm d t}}= \delta(t-t') \langle [A(t), B(t')] \rangle +
\langle \langle [ A(t),{\mathrm H}]; B(t') \rangle \rangle.
\eeqn 

The response or correlated functions, $\mathcal F_{AB}(t,t')$ and $\mathcal F_{BA}(t,t')$, are defined as the average over the statistical ensemble of the product of operators in the Heisenberg representation:

\beqn 
\mathcal F_{BA}(t,t')& = & \langle B(t') A(t) \rangle, \nonumber \\
\mathcal F_{AB}(t,t')& = & \langle A(t) B(t') \rangle.
\label{corre}
\eeqn 
Notice that Green Functions are a linear combination of the correlated functions given in Eqs. (\ref{corre}).

To proceed with the
calculation of the Green Functions it is important to know its spectral representations. 
The spectral intensity, $J(\om)$, is defined through the Fourier Transform of $\mathcal F_{AB}(t,t')$ by

\beqn 
J(\om) & = & \frac 1{2 \pi}~\int_{-\infty}^{\infty} 
{\mathcal F}_{AB}(t,t')~\re^{{\bf i} \om (t-t')} {{\rm d}(t-t')}, \nonumber\\
& = & \sum_{n,m}~\langle \overline \psi_n|B|\widetilde \phi_m \rangle
\langle \overline \psi_m|A|\widetilde \phi_n \rangle \re^{-\beta \widetilde E_n}~
\delta(\om-\widetilde E_n+\widetilde E_m).
\label{jw}
\eeqn 
In the same form, we introduce $J'(\om)$ as

\beqn 
J'(\om) & = & \frac 1{2 \pi}~\int_{-\infty}^{\infty} 
{\mathcal F}_{BA}(t,t')~\re^{{\bf i} \om (t-t')} {{\rm d}(t-t')}, 
\nonumber\\
& = & \re^{\beta \om}~\sum_{n,m}~\langle \overline \psi_n|B|\widetilde \phi_m \rangle
\langle \overline \psi_m|A|\widetilde \phi_n \rangle \re^{-\beta \widetilde E_n}~
\delta(\om-\widetilde E_n+\widetilde E_m),\nonumber \\ 
& = &  \re^{\beta \om}~J(\om).
\label{jwp}
\eeqn 

The spectral representation of $G_r(t,t')$ and $G_a(t,t')$ are given by their corresponding Fourier Transforms

\beqn 
G_{r,a}(E) & = & 
\frac 1{2 \pi}~\int_{-\infty}^{\infty} 
{G}_{r,a}(t,t')~\re^{{\bf i} E (t-t')} {{\rm d}(t-t')} 
\nonumber \\ 
& = & \frac 1{2 \pi}~\int_{-\infty}^{\infty} 
~(\re^{{\beta} \om }-1)J(w) \frac {{\rm d} \om}{E-\om \pm {\bf i} \epsilon}.
\eeqn 
Moreover, defining

\beqn 
G =  \frac 1{2 \pi}~\int_{-\infty}^{\infty} 
~(\re^{{\beta} \om }-1)J(\om) \frac {{\rm d} \om}{E-\om },
\eeqn 
it is straightforward to probe that

\beqn 
(\re^{{\beta} E }-1)J(E)= -{\bf i}~ & \rm{Lim} &  (G(E+{\bf i} \epsilon)-G(E-{\bf i} \epsilon)).
\nonumber \\
& \epsilon \rightarrow 0^+&
\eeqn  

Consider a system described by a Hamiltonian $\mathrm H_0$, which is perturbed by a time dependent interaction $\mathrm H_I$. Let the density matrix at $t=t_0$ be the equilibrium density matrix $\hat \rho_0= \frac 1 Z_0 \re^{-\beta H_0}$, with $Z_0={\mathrm{Tr}}(\hat \rho_0)$. To analyse the evolution in time of the system we have 
to evaluate $\rho(t)$ when $H=H_0+H_I(t)$ as:

\beqn 
 {\bf i} \hbar \frac{ {\rm d} \rho(t)}{\rm{dt}} & = & [ H_0+ H_I, \rho(t)],
\label{rhot} 
\eeqn 
with

\beqn 
\rho(t) & \rightarrow & \rho_0= \frac 1 Z \re^{-\beta H_0},
\eeqn
for $t \rightarrow t_0 $. 

Let us take $\rho(t)=\rho_0 + \Delta \rho(t)$, in linear the approximation

\beqn 
 {\bf i} \hbar \frac{ {\rm d} \Delta \rho(t)}{\rm{dt}} & \approx & [ H_0, \Delta \rho(t)]+ [ H_I, \rho_0],
\eeqn 
and $\Delta \rho(t) \rightarrow  0$ as $t \rightarrow t_0 $. 
So that

\beqn 
\Delta \rho(t) = -{\bf i} ~\int_{-\infty}^t ~\re^{{\bf i} H_0 (\tau-t)/\hbar} [H_I,\rho_0] \re^{-{\bf i} H_0 (\tau-t)/\hbar}~{\rm d} \tau
\label{delrho}
\eeqn 
Consequently, the mean value of an operator $\hat O$ is given by

\beqn 
\langle {\overline {\hat O(t)}} \rangle & = & \rm{Tr}(\hat O {\hat{\rho}}_0) + \rm{Tr} ( \hat O \Delta {\hat{\rho}}(t)), 
\nonumber \\
& = &  \langle \hat O \rangle_0  -{\bf i} ~\int_{-\infty}^t ~\langle [\hat O (t),H_I(\tau)] \rangle_0 ~{\rm d} \tau, \nonumber \\
& = &  \langle \hat O \rangle_0 + ~\int_{-\infty}^t ~( -{\bf i} \Theta(t-\tau) \langle [\hat O (t),H_I(\tau)] \rangle_0) ~{\rm d} \tau, \nonumber \\
& = &  \langle \hat O \rangle_0 + ~\int_{-\infty}^t ~\langle \langle \hat O (t);H_I(\tau) \rangle \rangle^r_0 ~{\rm d} \tau,
\label{ameanap}
\eeqn
where $Q(t)=U(t,t_0)^\dagger Q(t_0) U(t,t_0)$.

\section{Results and Discussion.}\label{results}

We shall apply the previous formalism to study the properties of Swanson's Hamiltonian \cite{swanson} at finite temperature. It reads

\begin{equation}
{\mathrm H}_{SW}^\times (\omega,\alpha,\gamma)= \hbar \omega~ \left( \aap \aam+ \frac 12 \right) +
\hbar \alpha~ {\aam}^2+  \hbar \gamma ~ {\aap}^2.
\label{hsw}
\end{equation}
In Eq.(\ref{hsw}), $\hbar \om$ fixes the energy scale. The parameter $\om$ stands for the frequency of a harmonic oscillator of mass $m_0$ and restoring parameter $k_0$, $\om=(k_0/m_0)^{1/2}$.
The coupling strengths of the Swanson model are given by $\hbar \al$ and $\hbar \gamma$, being $\al , \gamma \in \mathbb{R}$.

The adjoint Hamiltonian of ${\mathrm H}_c$ is
\begin{equation}
{\mathrm H}_c^\times (\omega,\alpha,\gamma)= \hbar \omega~ \left( \aap \aam+ \frac 12 \right) +
\hbar \gamma~ {\aam}^2+  \hbar \al ~ {\aap}^2.
\label{hswc}
\end{equation}

Following \cite{nos1}, we shall introduce the usual operators $\hat x$ and $\hat p$ as:

\beqn
 {\hat x} = \frac{b_0}{\sqrt{2}} \left ( \aap + \aam \right), ~~~~
 {\hat p} = \uni \frac{\hbar}{\sqrt{2} b_0} \left ( \aap - \aam \right),
\eeqn
so that, the Hamiltonian of (\ref{hsw}) can be written as

\beqn
{\mathrm H}_{SW}^\times (\omega,\alpha,\gamma) & = &
\frac 1 2 \hbar (\omega + \alpha + \gamma ) \left( \frac {\hat x} {b_0}\right)^2 
+\frac 1 2 \hbar (\omega - \alpha - \gamma ) \left( \frac {b_0~ \hat{p}} \hbar \right)^2
\nonumber \\
& &+  \hbar \frac{(\alpha-\gamma)}{2} \left( 2~\hat{x} \frac{ {\bf i}}{\hbar} \hat{p}+1 \right).
\label{hxp}
\eeqn

In absence of a time-dependent perturbation, the Swanson Hamiltonian,  for $\om-\al-\gamma \neq 0$, can be mapped to a self-adjoint Hamiltonian, $\mathfrak h_{SW}^\times$, through  a gauge transformation given by:
\beqn
\Upsilon= \re^{-\frac{\al-\gamma}{2(\om-\al-\gamma)} \hat x^2}.
\label{ups0}
\eeqn 
Thus
\beqn
\Upsilon~ {\mathrm H} ~\Upsilon^{-1} ={\mathfrak h}_{SW}, \\
\label{simi1}
\Upsilon^{-1}~{\mathrm H}_c ~\Upsilon  = {\mathfrak h}_{SW}.
\label{simic}
\eeqn
The explicit form of the Hamiltonian  ${\mathfrak h}_{SW}$ is
\beqn
{\mathfrak h}_{SW}^\times = \frac 1 {2 m} \left ( \hat p^2+ m^2 \Omega^2 \hat x^2 \right ),
\label{hosc}
\eeqn
with 

\beqn
m&=&\frac{ \hbar}{(\omega-\alpha-\gamma) b_0^2},\nonumber \\
\Omega & = & \sqrt{\om^2- 4 \al \gamma}.
\label{masseff}
\eeqn

In a previous work \cite{nos1}, we have shown that the Hamiltonian of Eq.(\ref{hsw}) modelled different physical problems depending on the values adopted by its parameters.
There are four possible regions in the parameter model space. These regions are characterised by $\Omega= |\Omega| \re^{\bf i \phi}$ and the effective parameter $m$. 

Following the notation of \cite{nos1}, Region I corresponds to the sector of model space for which both $m$ and $\Omega^2$ take positive values. In Region I the Hamiltonian is similar to the usual harmonic oscillator.  The case $m>0$ and $\Omega^2<0$ corresponds to an inverted oscillator, Region II \cite{maru,chu1,chu2,iho0,iho1,iho2,iho3,iho-chaos0,iho-chaos}. If the $m<0$ and $\Omega^2>0$ can be interpreted as a harmonic oscillator with negative mass, Region III \cite{glauber,neg-mass-1,neg-mass-2,neg-mass-3,neg-mass-4,neg-mass-5}. Finally, the case $m<0$ and $\Omega^2<0$ can be interpreted as a parabolic barrier for a system with negative mass, Region IV. 

In what follows, we shall take ${\mathrm H}_0={\mathrm H}_{SW}$. 

We shall measure $k_B T$ and $\hbar \Om$ in the same arbitrary energy unit. Consequently, $T$ will be given in units of $[energy/k_B]$ and $\Om$ in units of $[energy/\hbar]$. 

\subsection{Example 1.} \label{ej1}

As a first example, we shall consider a perturbation of the form 

\beqn 
{\rm H}_I (t) & =&  v(t) \hat x^2, 
\eeqn
with $v(t)=-V \cos (W t) \re^{\epsilon t}$, with $\epsilon>0$, with $W$ and $\epsilon$ given in units of $[energy/\hbar]$, and 
$V$  in units of energy. We shall adopt parameters $\alpha,~\gamma$ and $\omega$ consistent with the Region I of the model space. In Region I, the eigenfunctions of $H_{SM}$ belong to the usual Hilbert space.  The Hamiltonian reads

\beqn   
{\rm H}={\rm H}_{SM}+{\rm H}_I(t).
\eeqn   
In this case, we can choose an $\Upsilon$ independent of $t$. In particular, for   $\Upsilon$ of Eq. (\ref{ups0}), ${\rm H}$ is similar to the Hamiltonian ${\mathfrak h}$ given by
\beqn  
{\mathfrak h} & = &  {\mathfrak h}_0+{\mathfrak h}_I,\nonumber \\
{\mathfrak h}_0 & = &  \frac 1 {2 m} \left ( \hat p^2+ m^2 \Omega^2 \hat x^2 \right ), \nonumber \\
{\mathfrak h}_I & = & v(t) \hat x^2.
\label{hoscej1}
\eeqn

The approximate solution can be obtained from Eq.(\ref{ameanap}). It should be mention that the linear approximation is  valid for $V << \Om ^2$. In Appendix B we have resumed the principal steps in order to obtain:

\beqn 
\langle \overline{\hat x^2(t,\beta)} \rangle & = & 
\tilde b_0^2 \coth \left(\frac {\beta \hbar \Omega}{2} \right) 
\left (\frac 1 2  + V ~ I_1 \right),  \nonumber \\
\langle \overline{\hat p^2(t,\beta)} \rangle
& = &\frac{\hbar^2}{\tilde b_0^2}
\coth \left(\frac {\beta \hbar \Omega}{2} \right)\left (\frac 1 2 - V ~ I_1 \right),
\eeqn
with

\beqn 
I_1& = & \int_{-\infty}^t ~\cos(W \tau) \re^{\epsilon \tau} \sin(2 \Omega (t-\tau)) {\rm d} \tau \nonumber \\
& = & -2 \Omega \re^{t \epsilon }~
\frac {[(W+2\Omega)(W-2\Omega)-\epsilon^2]~\cos (Wt)- 2 W \epsilon ~ \sin (W t) }
{ [(W+2 \Omega )^2+\epsilon ^2][(W-2\Omega )^2+\epsilon ^2] }. \nonumber \\
\eeqn 

To estimate the benefits of the method we shall compare the approximate results with the exact ones.

To compute the exact result, we shall write the time evolution as $U(t,t_0)= U_0(t,t_0) U_I(t,t_0)$, being $U_0(t,t_0)=\re^{-{\bf i} {\mathfrak h}_0 (t-t_0)}$. The equation to solve, in order to determine $U_I(t,t_0)$, can be obtained straightforward and it reads

\beqn 
{\bf i} \hbar \frac {\partial U_I}{\partial t}= v(t) ~ \hat x^2(t)~U_I, 
\eeqn
with $\hat x^2(t)= U_0^{-1}(t,t_0) \hat x^2 U_0(t,t_0)$ and satisfying 
the initial the condition $U_I(t) \rightarrow  I$ for $t  \rightarrow -\infty$. 

As we have assumed that the parameters $\alpha$ and $\beta$ correspond to  Region I, the spectrum is real and discrete. To carry out the calculation it is convenient to introduce

\beqn
\aamt & = & \frac{1}{\sqrt{2}} \left ( \frac {\hat x} {\tilde b_0} + {\bf i} \frac {\tilde b_0} {\hbar} \hat{p}\right), \nonumber \\
\aapt & = & \frac{1}{\sqrt{2}} \left ( \frac {\hat x} {\tilde b_0} - {\bf i} \frac {\tilde b_0} {\hbar} \hat{p}\right), \nonumber \\
\eeqn
we have defined $\tilde b_0$ as the characteristic length of the harmonic oscillator of Eq. (\ref{hosc}), that is
$\tilde b_0=\sqrt{\hbar/ (m \Om)}$.

In the previous equation,
$\hat x^2(t)= \frac {{\tilde b}_0^2} {2} \left({\tilde a}^{\dagger 2} \re^{{\bf i} 2 \Omega t} + {\aamt}^2 \re^{-{\bf i} 2 \Omega t }+2 {\tilde a^\dagger} \tilde a  +1\right)$.
After some algebra we can write the solution as

\beqn 
U_I(t) = \re^{-{\bf i} \left( \kappa \frac {{\tilde a}^{\dagger 2}}2 + \kappa^* \frac{{\aamt}^2}2 + \kappa_0 ({\aapt} \aamt  +\frac 12 )\right)},
\eeqn 
where

\beqn 
\kappa   & = & -V \re^{\epsilon t} \re^{{\bf i} 2 \Omega t} 
\frac {(W^2+(\epsilon-2 {\bf i} \Omega)^2)((\epsilon+2 {\bf i} \Omega) \cos(W t) + W \sin(W t))}
{(\hbar (W-2  \Omega)^2 +\epsilon^2)(W+2  \Omega)^2 +\epsilon^2))}, \nonumber \\
\kappa_0 & = &  -V \re^{\epsilon t} \frac {\epsilon \cos(W t) + W \sin(W t)}
{\hbar (W^2 +\epsilon^2)}.
\eeqn 

\begin{figure}
\includegraphics[width=\textwidth]{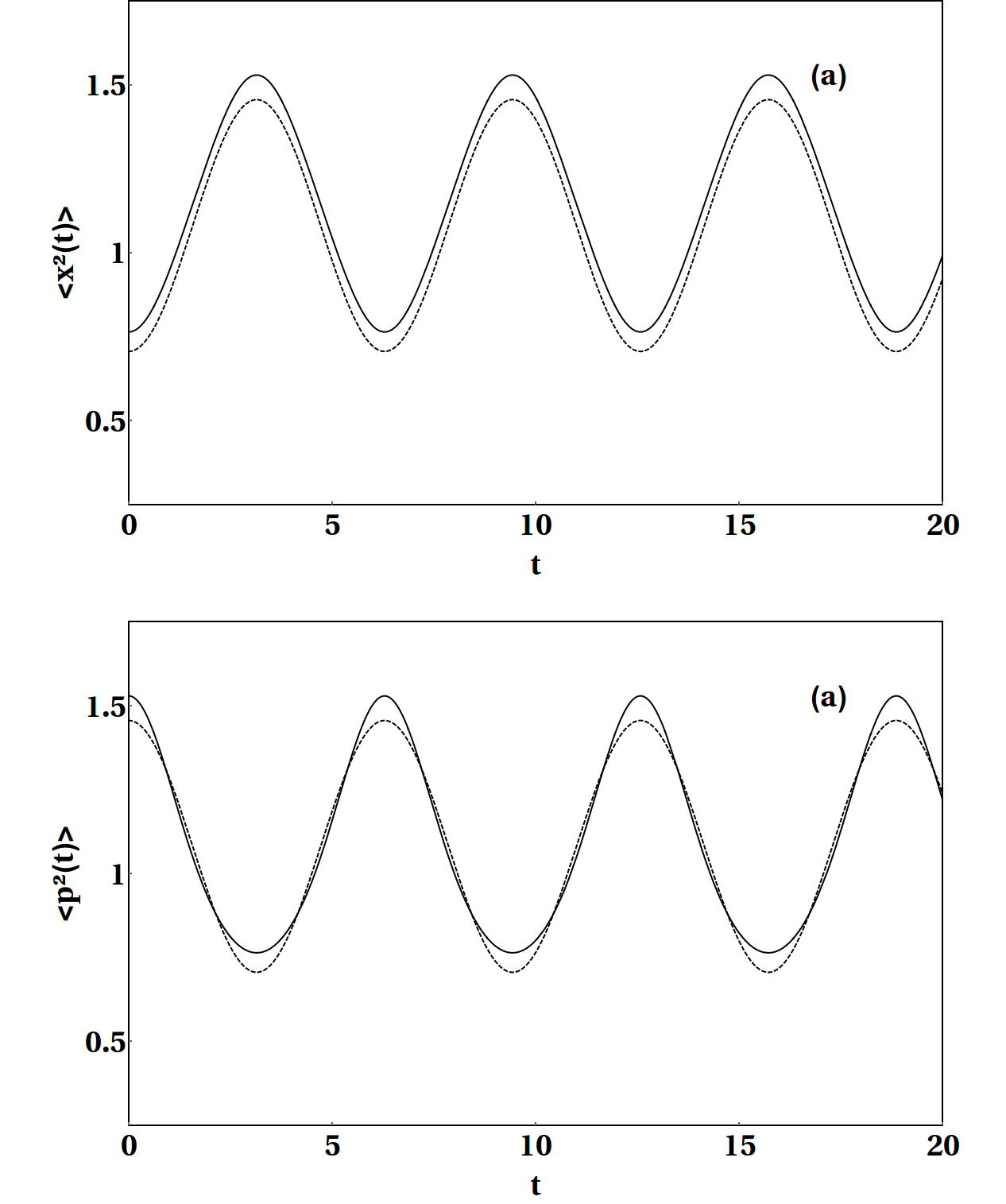}
\caption {The behaviour of $\langle{\overline{\hat x^2(t)}} \rangle$ and $\langle{\overline{\hat p^2(t)}}\rangle$ as a function of the time $t$ is depicted in Panels (a) and (b), respectively. We have plotted the exact and approximate results with solid and dashed curves, respectively. We have fixed the value of the parameters to $V=0.26$, $\Om=W=1$ and $\beta=1$. }\label{fig1}
\end{figure}

\begin{figure}
\includegraphics[width=\textwidth]{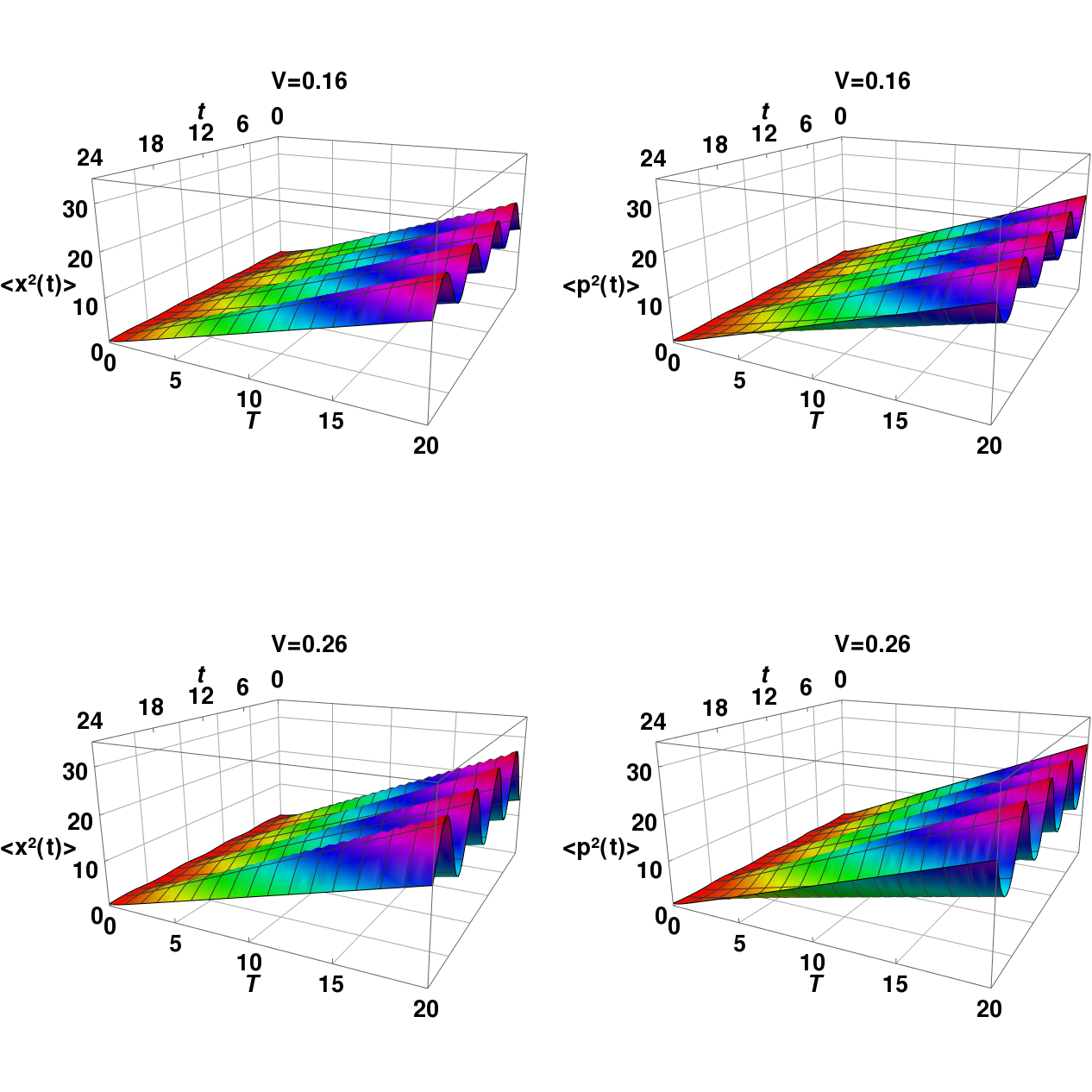}
\caption {The behaviour of $\langle{\overline{\hat x^2(t)}}\rangle$ and $\langle{\overline{\hat p^2(t)}}\rangle$ as a function of the time $t$ and the temperature $T$. 
The results for $\langle{\overline{\hat x^2(t)}}\rangle$ are drawn in Panels (a) and (c), while those for $\langle{\overline{\hat p^2(t)}}\rangle$ are plotted in Panels (b) and (d). In Panels (a) and (b),  $V$ was fixed to the value 0.16, while in Panels (c) and  (d), $V=0.26$.
We have fixed $\Om=W=1$. }\label{fig2}
\end{figure}

Having obtained $U(t)$, we can solve the equation for $\rho(t)$. If we propose $\rho(t)=\re^{ -{\bf i} H_0 t} \rho_I(t) \re^{ {\bf i} H_0 t}$, Eq.(\ref{rhot}) reads

\beqn 
 {\bf i} \hbar \frac{ {\rm d} \rho_I(t)}{\rm{dt}} & = & [ {\mathfrak h}_I, \rho_I(t)],
\label{rhotI} 
\eeqn 
and its solution is given by $\rho_I(t)= U_I(t) \rho_0 U_I^\dagger$,
with the initial condition $\rho_0=\re^{-\beta {\mathfrak h}_0}/Z$.

For a given operator $\hat A$, we aim to obtained

\beqn 
\langle {\overline{ A(t)}} \rangle=\frac 1 Z {\rm Tr} (\rho_0 U(t)^{-1} A U(t)).
\eeqn

For ${\mathfrak h}_0= \hbar \Omega \left( a^\dagger a + \frac 12 \right)$, we have

\beqn 
Z= {\rm Tr}({\re^{-\beta {\mathfrak h}_0}})=\frac {\re^{- \frac 12 \hbar \Omega \beta}}{1-\re^{-\hbar \Omega \beta }}.
\eeqn

Following the steps pointed out in Appendix C:

\beqn 
\langle {\overline {\hat x^2(t)}} \rangle & = & \frac{\tilde b_0^2}{~2~}\coth \left(\frac{\beta  \Omega }{2}\right)
   \left( 1-2 \frac {\zeta_+ \zeta_-}{\zeta_0^2} +
   \left ( \re^{2 {\bf i} \Omega t} \frac{\zeta_-} {\zeta_0^2}+
           \re^{-2 {\bf i} \Omega t} \frac{\zeta_-^*} {{\zeta_0^*}^2}
   \right ) 
   \right),
\nonumber \\
\langle {\overline {\hat p^2(t)}} \rangle &=& \frac{\hbar^2}{2 \tilde b_0^2} \coth \left(\frac{\beta  \Omega }{2}\right)
   \left( 1-2 \frac {\zeta_+ \zeta_-}{\zeta_0^2} -
   \left ( \re^{2 {\bf i} \Omega t} \frac{\zeta_-} {\zeta_0^2}+
           \re^{-2 {\bf i} \Omega t} \frac{\zeta_-^*} {{\zeta_0^*}^2}
   \right ) 
   \right).
   \nonumber \\
\label{amean}
\eeqn

In the previous equation we have defined the quantities $\zeta_\pm$ and $\zeta_0$ as

\beqn 
\zeta_0 & = & \left ( \cos(d)+ {\bf i} \frac {\kappa_0} d (\sin(d) \right )^{-1}, \nonumber \\
\zeta_+ & = & -{\bf i} \frac {\kappa} d \sin(d) \zeta_0 , \nonumber \\
\zeta_- & = & -{\bf i} \frac {\kappa^*} d \sin(d) \zeta_0 , \nonumber \\
\eeqn  
with $d=\sqrt{\kappa_0^2-|\kappa|^2}$.

In Figure \ref{fig1} we plot the exact and approximate solutions for $\langle\overline{\hat x^2(t,\beta)}\rangle$ and $\langle\overline{\hat p^2(t,\beta)}\rangle$ as a function of time. Panel (a) and (b) correspond to the results we have obtained for $\langle\overline{\hat x^2(t,\beta)}\rangle$ and $\langle \overline{\hat p^2(t,\beta)}\rangle$, respectively. We have plotted the exact and approximate results with solid and dashed curves, respectively. We have fixed the value of the parameters to $V=0.26$, $\Om=W=1$ and $\beta=1$. As seen from the Figure, approximate solutions follow the trends of the exact ones with high accuracy. 

The behaviour of $\langle {\overline{\hat x^2(t)}}\rangle$ and ${\overline{\hat p^2(t)}}\rangle$ as a function of the time $t$ and the temperature $T$ is plotted in Figure \ref{fig2}. 
The results for $\langle{\overline{\hat x^2(t)}}\rangle$ are drawn in Panels (a) and (c), while those for $\langle{\overline{\hat p^2(t)}}\rangle$ are plotted in Panels (b) and (d). In Panels (a) and (b),  $V$ was fixed to the value 0.16, while in Panels (c) and  (d), $V=0.26$.
We have fixed $\Om=W=1$. As is observed from Figure \ref{fig2}, the periodic pattern in time persists as the temperature is increased.



\subsection{Example 2.} \label{ej2}

In this example, we take values of the parameters, $\alpha,~\gamma$ and $\omega$, which are compatible with Region I. Next, we assume that at $t=t_0$ an interaction term ${\rm H}_I(t)=- v(t) ~ \left( \hat x \hat p+ \hat p \hat x \right)$ is turned on, with a periodic time dependence,  $v(t)$, given by 
\beqn
v(t)=\frac 12 V \left( \alpha-\gamma + 
{\bf i}  \frac {{\rm d} f(t)}{{\rm d} t} \right),
\eeqn
with

\beqn
f(t)= \ln (g(t)),~~~g(t)= M_C \left (\frac{4 (\alpha-\gamma)^2}{W^2},\frac {2 V}{W^2},\frac {W t} 2\right),
\eeqn 
$M_C$ stands for the even solutions of Mathieu's equation.
The values of $W$ are given in units of $[energy/\hbar]$, and 
$V$  in units of energy.

In the present case, the operator $\Upsilon$, which was introduced in Eq. (\ref{ht}), depends on the time. We have chosen the operator $\Upsilon$ as

\beqn 
\Upsilon(t)= \re^{-\frac{\al-\gamma- 2 v(t) }{2(\om-\al-\gamma)} \hat x^2}.
\eeqn

After some algebra, we obtained ${\mathfrak h}$ of Eq. (\ref{ht}). It reads

\beqn
{\mathfrak h}= \frac{1}{2 m} \hat p^2+ \frac 12 m^2 \Omega^2 \hat x^2 -V \cos(W t) \hat x^2, ~~~ t \ge t_0.
\eeqn 

In the approximate case, for $t \ge t_0$, we finally obtain

\beqn 
\langle \overline{\hat x^2(t,\beta)} \rangle & = & 
\tilde b_0^2 \coth \left(\frac {\beta \hbar \Omega}{2} \right) 
\left (\frac 1 2  + V ~ {\rm I_2} \right),  \nonumber \\
\langle \overline{\hat p^2(t,\beta)} \rangle
& = &\frac{\hbar^2}{\tilde b_0^2}
\coth \left(\frac {\beta \hbar \Omega}{2} \right)\left (\frac 1 2 - V ~ {\rm I_2} \right).
\eeqn

\beqn 
{\rm I_2} & = & \int_{t_0}^t ~V(\tau) \sin(2 \Omega (t-\tau)){\rm d} \tau\nonumber \\
& = &-\frac{ 1}{\left(W^2 -4\Omega ^2\right)}
\left[
2 \Omega ~(\cos(Wt)-\cos(Wt_0)~\cos(2 \Omega (t-t_0)))+ \right . \nonumber \\
&& ~~~~~~~~~~~~~~~~~~~~~~~~~~~~ \left . W~\sin(Wt_0)~\sin(2 \Omega (t-t_0)))
\right ]
.\nonumber \\
\eeqn 
\begin{figure}
\includegraphics[width=\textwidth]{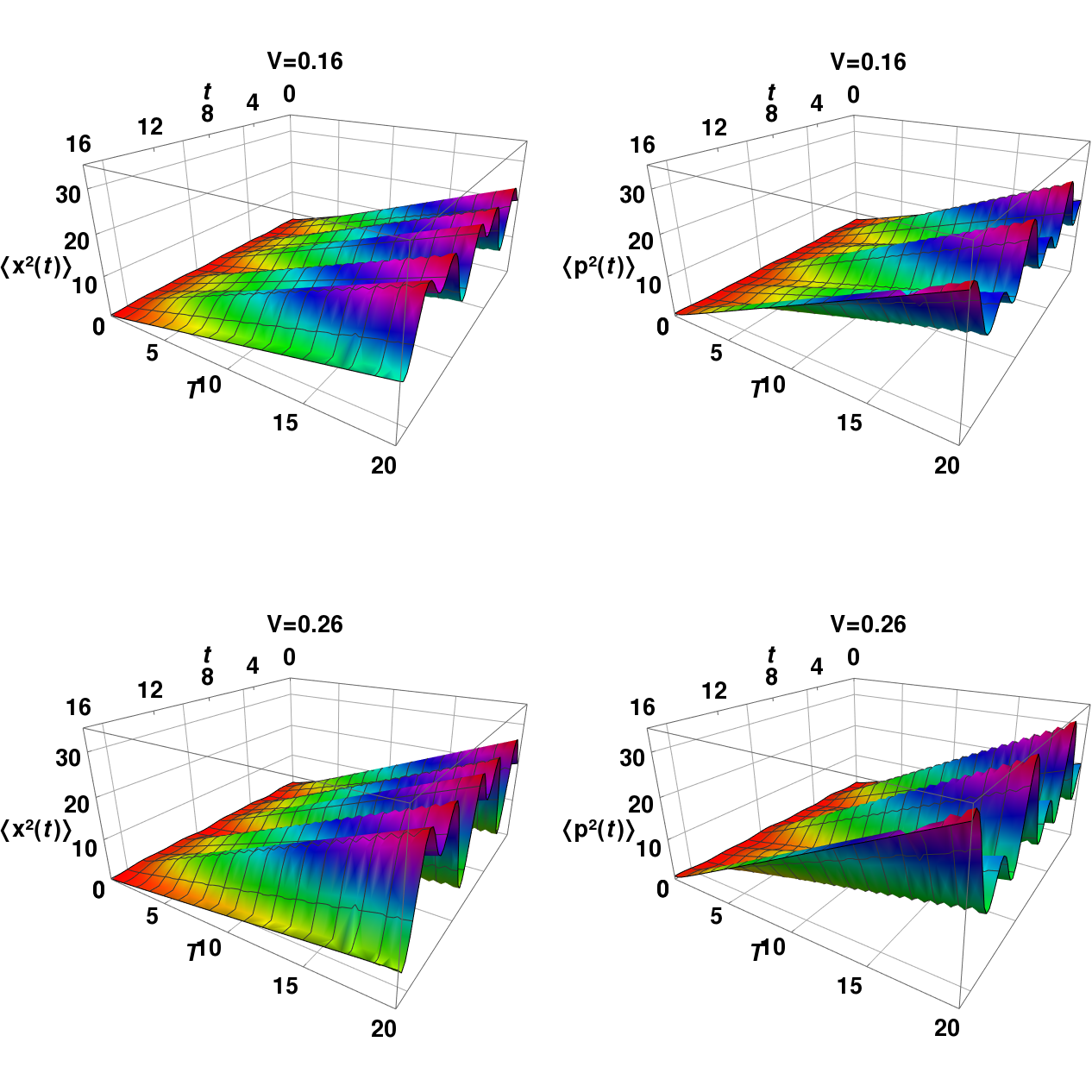}
\caption {The behaviour of ${\overline{\hat x^2(t)}}$ and ${\overline{\hat p^2(t)}}$ as a function of the time $t$ and the temperature $T$. 
The results for ${\overline{\hat x^2(t)}}$ are drawn in Panels (a) and (c), while those for ${\overline{\hat p^2(t)}}$ are plotted in Panels (b) and (d). In Panels (a) and (b),  $V$ was fixed to the value 0.16, while in Panels (c) and  (d), $V=0.26$.
We have fixed $\Om=W=1$. }\label{fig3}
\end{figure}

In Figure \ref{fig3}, we have plotted the behaviour of ${\overline{\hat x^2(t)}}$ and ${\overline{\hat p^2(t)}}$ as a function of the time $t$ and the temperature $T$, for two values of the coupling constant $V$. As commented in the previous example, the perturbative approach is valid for $V << \Om$. The results for ${\overline{\hat x^2(t)}}$ are drawn in Panels (a) and (c), while those for ${\overline{\hat p^2(t)}}$ are plotted in Panels (b) and (d). In Panels (a) and (b),  $V$ was fixed to the value 0.16, while in Panels (c) and  (d), $V=0.26$.
We have fixed $\Om=W=1$. The periodic behaviour in time is preserved as the temperature is increased. 

\subsection{Example 3.}\label{ej3}

As a final example, we shall take values of $\alpha$, $\beta$ and $\omega$ consistent with the broken PT-symmetry phase, Region II. We have shown in \cite{nos1} that, in Region II, the Swanson Hamiltonian can be mapped to an inverted oscillator ${\mathfrak h}^\times$, of positive mass and frequency $\Om^2<0$:

\beqn 
{\mathfrak h}_0^\times = \frac{\hat p^2}{2 m}-\frac 12 m |\Om|^2 \hat x^2.
\label{iho}
\eeqn 
We shall study the effect of applying a perturbation to the potential of the form $v(t)= V \cos(W t) \hat x^2 \Theta(t-t_0)$. 
The values of $W$ are given in units of $[energy/\hbar]$, and 
$V$  in units of energy. 

Let us first review the physics associated with the Hamiltonian ${\mathfrak h}_0^\times$ of Eq. (\ref{iho}).
Following the work \cite{maru}, after transformation to new variables $\hat u$ and $\hat v$, ${\mathfrak h}_0^\times$ can be written as

\beqn 
{\mathfrak{h}}_0^\times = - \frac 1 2 \hbar |\Om| \left(  \hat{u} \hat{v} +  \hat{v} \hat{u} \right),
\eeqn 
with
\beqn
{\hat u} = \sqrt{\frac{m |\Om|}{2 \hbar}  } \left( \hat{x}- \frac{1}{m |\Om|} {\hat p} \right),
~~~~{\hat v} = \sqrt{\frac{m |\Om|}{2 \hbar}  } \left( \hat{x}+ \frac{1}{m |\Om|} {\hat p} \right).
\label{uv2}
\eeqn
The characteristic scale of length is defined as $\tilde b_0 = \sqrt{\hbar/ (m |\Om|)}$.
The discrete eigenvalues and the corresponding generalised eigenvectors can be summarised as follows,

\beqn
\begin{array}{cl}
     \hv |\phi_0^+ \rangle =0, & \varepsilon_0^+=E_0 \\
     |\phi_n^+ \rangle=\frac {1 }{\sqrt{n!}} {\hat u}^{ n} |\phi_0^+ \rangle, & \varepsilon_n^+= E_n ,
\end{array}
\eeqn
and
\beqn
\begin{array}{cl}
     \hu |\phi_0^- \rangle =0, & \varepsilon_0^-=-E_0 \\
     |\phi_n^- \rangle=\frac {(-{\bf i})^n}{\sqrt{n!}} {\hat v}^{ n}|\phi_0^- \rangle, & \varepsilon_n^-=-E_n,
\end{array}
\eeqn
where $E_n= \uni   \frac{\hbar |\Om|} 2 (n+1/2)$.  
Also, it is straightforward to prove that

\beqn
\langle \phi_m^\mp | \phi_n^\pm \rangle= \langle \phi_m^\pm | \phi_n^\pm \rangle_S = \delta_{mn},
\label{qort0}
\eeqn
being $S=\sum_n ~(|\phi_n^- \rangle \langle \phi_n^- |+|\phi_n^+ \rangle \langle \phi_n^+ |)$. In writing Eq.(\ref{qort0}), we have used the fact that ${\mathfrak{h}}^\times$ is an auto-adjoint operator with discrete complex eigenvalues \cite{rhs-1}.

We can propose for the spectral representation of ${\mathfrak h}_0^\times$ the following decomposition

\beqn 
{\mathfrak h}_0^\times & = & {\mathfrak h}_+ + {\mathfrak h}_-, \nonumber \\
{\mathfrak h}_+ & = & \sum_n ~E_n   |\phi_n^+ \rangle  \langle \phi_n^-| \nonumber \\
{\mathfrak h}_- & = & \sum_n ~E_n^* |\phi_n^- \rangle  \langle \phi_n^+|.
\eeqn
Observed that ${\mathfrak h}_+^\dagger={\mathfrak h}_-$.

As reported in \cite{maru}, the temporal evolution operator $U(t)^\times=\re^{{\bf i} {\mathfrak h}^\times t}$ is not unitary on $\Phi^\times$, i.e. it does not preserve the norm of the states $|\phi_m^\pm \rangle$, 

\beqn 
U(t)^\times | \phi_n^+ \rangle & = &
\re^{ \uni  h_+ t} | \phi_n^+ \rangle =
\re^{-\hbar |\Om|(n+1/2)~t} | \phi_n^+ \rangle, ~t \ge 0, \nonumber \\
U(t)^\times | \phi_n^- \rangle & = & 
\re^{ \uni  h_- t} | \phi_n^- \rangle =
\re^{~\hbar |\Om|(n+1/2)~t} | \phi_n^- \rangle , ~t \leq 0.
\eeqn
To ensure an irreversible quantum mechanics formalism, we proceed as done in \cite{maru,rhs-tasy}. We define two new subspaces $\Phi^\pm$, such that $\Phi=\Phi_++\Phi_-$ ($\Phi_+ \cap \Phi_- \neq 0$ in general). Now, we construct two Gel'fand triplets \cite{rhs-1,hardy}, namely

\beqn
& & \Phi_+ \subset {\cal H} \subset \Phi_+^\times, \nonumber \\
& & \Phi_- \subset {\cal H} \subset \Phi_-^\times.
\label{gelfandpm}
\eeqn
At $t=t_0$,  $|\phi^+\rangle \in \Phi_+$ and $|\phi^-\rangle \in \Phi_-$. The operator $U(t)^\times$ is well defined on $\Phi_+$ for $t>0$, and it is well defined in $\Phi_-$ for $t<0$.

We shall, now, turn on the perturbative interaction term. That is 

\beqn    
{\rm H}^\times = {\rm H}_{SW}^\times+ {\rm H}_I^\times(t), ~~~{\rm H}_I^\times(t)=  V \cos(W t) \Theta(t-t_0)~{\hat x}^2.
\eeqn   
By choosing $\Upsilon$ of Eq. (\ref{ups0}) we can mapped the Hamiltonian $\rm H$ to 

\beqn    
{\mathfrak h}^\times = {\mathfrak h}_{0}^\times+ {\mathfrak h}_I^\times(t), ~~~{\mathfrak h}_I^\times(t)=  V \cos(W t) \Theta(t-t_0)~{\hat x}^2.
\eeqn   

As we want to study the time evolution for $t>0$ \cite{rhs-tasy}, in computing the density matrix operator, we shall trace the contribution of the anti-resonant states.  

The expansion of $\rho(t)$ is given by 

\beqn 
& & \rho(t) \approx \rho_0
- \frac {\bf i}{\hbar}~\re^{{\bf i} {\mathfrak h}_0^\times t/\hbar} ~\int_{-\infty}^t {\rm d} \tau  ~ 
[{\mathfrak h}_I^\times(\tau),\rho_0] 
~\re^{-{\bf i} {\mathfrak h}_0^\times t/\hbar}
\nonumber \\
& & ~~~+
\left( \frac {\bf i}{\hbar} \right)^2 \re^{{\bf i} {\mathfrak h}_0^\times t/\hbar} ~
\int_{-\infty}^t ~{\rm d} \tau~\int_{-\infty}^\tau ~{\rm d} \tau'
[{\mathfrak h}_I^\times(\tau),[{\mathfrak h}_I^\times(\tau'),\rho_0]] \re^{-{\bf i}~ {\mathfrak h}_0^\times t/\hbar}+...
\nonumber \\
\eeqn 
Notice that the lower order term that contributes to the density matrix, $\rho(t)$, is the term of second order.

In Appendix D and Appendix E, we have summarised the steps of the calculation. After some cumbersome algebra, the reduced density matrix, $\rho_r(t)$, can be written as
\beqn  
\rho_r(t) & \approx & \rho_r^{(0)} + \rho_r^{(2)}, \nonumber\\
\rho_r^{(0)} & = & \frac{\re^{-\beta {\mathfrak h}_+}}{Z_r}, ~~~Z_r= \frac{1}{\uni 2 \sin(\beta |\Om|/2)}, \nonumber \\
\rho_r^{(2)} & = & 
-\re^{\uni {\mathfrak h}_+^\dagger t} \frac{\re^{-\beta {\mathfrak h}_+}}{Z} \re^{-\uni {\mathfrak h}_+ t}~{\mathfrak F}(t,\beta), ~~~Z= \frac{1}{( 2 \sin(\beta |\Om|/2))^2}, \nonumber \\
\eeqn 
where ${\mathfrak F}(t,\beta)$ can be expressed as

\beqn  
{\mathfrak F}(t,\beta) & = & {\rm I}_1  ( |\eta|^2 S_+(\beta+ 2\uni  t)-2 \eta^*S_0(\beta+2\uni  t))+ \nonumber \\
& & {\rm I}_2  ( |\eta |^2 S_-(\beta+2 \uni t)+2 \eta S_0(\beta+2 \uni  t)).
\eeqn  
with
\beqn  
{\rm I}_1 & = & \frac{V^2}{2} \left( 
\frac{\sin(W t)^2}{W^2+ 4 |\Om|^2}-|\Om| \frac{2 W t +\sin(2 W t)}{W(W^2+ 4|\Om|^2)}+ \right .
\nonumber \\ 
& & 
\left. ~~~~ 4 |\Om| 
\frac{ \re^{2 |\Om| t} ( W \sin(W t) +2 |\Om| \cos(W t) )-2 |\Om|}{(W^2+ 4 |\Om|^2)^2}
\right), \nonumber \\
{\rm I}_2 & = & \frac{V^2}{2} \left( 
\frac{\sin(W t)^2}{W^2+ 4 |\Om|^2}-|\Om| \frac{2 W t +\sin(2 W t)}{W(W^2+ 4|\Om|^2)}- \right .
\nonumber \\ 
& & 
\left. ~~~~ 4 |\Om| 
\frac{ \re^{-2 |\Om| t} ( W \sin(W t)-2 |\Om| \cos(W t))+2 |\Om|}{(W^2+ 4 |\Om|^2)^2}
\right), \nonumber \\
S_-(y) &=&  \frac{\re^{\uni |\Om| \beta}}{4 \uni (\sin(y |\Om|/2))^3}, ~~~~
S_+(y) =  \frac{\re^{-\uni |\Om| \beta }}{4 \uni (\sin(y |\Om|/2))^3},\nonumber \\
S_0(y) &=& -\frac{\cos(y |\Om|)}{2 (\sin(y |\Om|/2))^2}, \nonumber \\
\eta & = & 2 \uni \re^{\uni \beta |\Om|} \sin(\beta |\Om|).
\eeqn   

It should be remarked that ${\mathfrak h}_+$ is not an auto-adjoint operator, ${\mathfrak h}_+^\dagger={\mathfrak h}_-$. 

In Appendix D we have derived the expression for $\rho_r^{(0)}$ and $Z_r$. From $Z_r$ we analyse the behaviour of the mean value of particles, $n_r^0(T,\Om)$, and the internal energy, $U_r^0(T,\Om)$. Alternatively, we can obtain the same results by 

\beqn  
n_r^{(0)}(T) & = & {\rm{Tr}} \left( -\frac 12 \{\hat u ,\hat v \} \rho_r^{(0)} \right )_S=
\frac 12 {\rm{cotg}} \left ( \frac 12 \hbar |\Om| \beta \right), \nonumber \\
U_r^{(0)}(T) & = & {\rm{Tr}} \left ( -\hbar |\Om| \frac 12 \{\hat u ,\hat v \} \rho_r^{(0)} \right)_S=
\frac 12 \hbar |\Om| {\rm{cotg}} \left( \frac 12 \hbar |\Om| \beta \right), \nonumber \\
C_r^{(0)}(T) & = & \frac{{\rm d} U_r}{{\rm d }T}=\left( \frac {\hbar |\Om|}{k_B T ~\sin(\hbar |\Om|/(2 T))} \right)^2.
\label{U0}
\eeqn 

The previous results suggest that there exists a lower bound to the temperature of the system, so that $k_B T \ge \hbar |\Om|/ \pi$. This conjecture is in agreement with the results advanced in \cite{maldacena,bh2,iho-chaos0}. That is, 
the Authors of \cite{maldacena} predict a bound to the rate of growth of chaos in thermal quantum
systems. They conjecture that chaos can develop no faster than in an exponential form, being the Lyapunov exponent, $\lambda_L \le 2 \pi k_B T/\hbar$. In a recent work 
\cite{iho-chaos0},
the quantum Lyapunov exponent, computed through the out-of-time-order correlator, for an inverted harmonic oscillator in one dimension has been derived. Moreover, in \cite{bh2} the Authors infer, taking into account the results of \cite{maldacena}, 
the existence of a lower bound to the temperature for a thermal quantum system with a fixed positive Lyapunov exponent. It should be observed that the Lyapunov exponent of the IHO of Eq. (\ref{iho}) is $\lambda_L=2 |\Om|$.

\begin{figure}
\includegraphics[width=\textwidth]{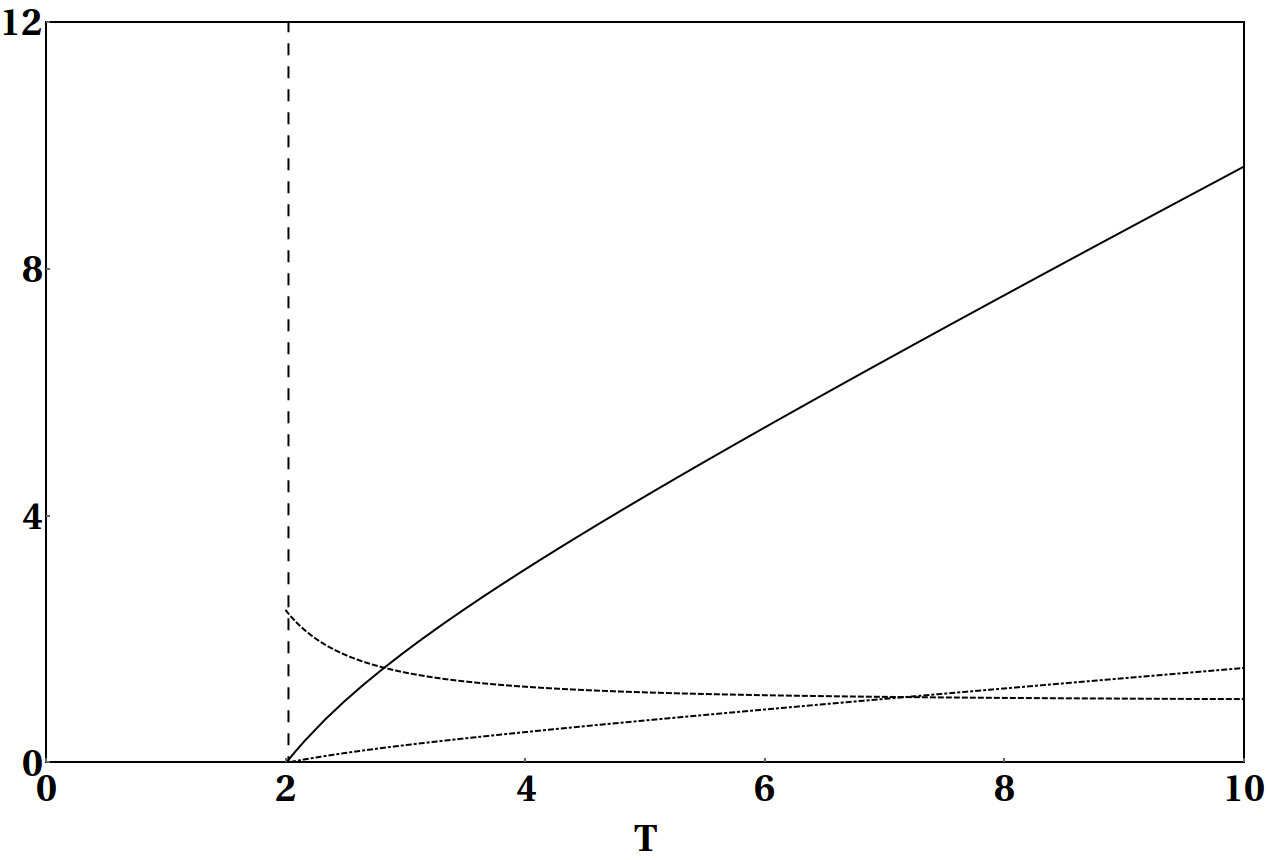}
\caption {The internal energy, 
$U_r^0(T,\Om)$, the specific heat, $C_r^0(T,\Om)$, and the mean value of the number of particles, $n_r^{(0)}(T)$, is depicted as a function of temperature $T$
For fixed values of the parameters $|\Om|= 2 \pi$, $W=\pi$ and $V=\pi$.
$U_r^{(0)}(T)$, $n_r^{(0)}(T)$ and $c_r^{(0)}(T)$ are depicted with solid, dashed and dotted-dashed curves, respectively.}\label{fig4}
\end{figure}

In Figure \ref{fig4}, we have plotted the internal energy, $U_r^0(T,\Om)$, the specific heat, $C_r^0(T,\Om)$, and the mean value of the number of particles, $n_r^{(0)}(T)$, as a function of temperature $T$, for $\hbar \Om=2 \pi$. $U_r^{(0)}(T)$, $n_r^{(0)}(T)$ and $c_r^{(0)}(T)$ are depicted with solid, dashed and dotted-dashed curves, respectively. As observed from the Figure, for large values of the temperature, $U_r^{(0)}(T)$ is linear in $T$ and, consequently, the values of the specific heat tend to be constant. 

If we switch on the periodic perturbation term and compute mean values associated with the number of particles to second order in the expansion of $\rho_r(t)$ can be expressed as 

\beqn  
\overline{n_r(t,\beta)} & = & \frac 12 {\rm{cotg} }(|\Om| \beta/2)- 
\frac{1}{2 Z} S_0(\beta-2 \uni t) {\mathfrak F}(t,\beta), \nonumber \\
\overline{U_r(t,\beta)} & = & \hbar |\Om| n_r(t,\beta).
\label{nt}
\eeqn   
In \label{a4} we have pointed out the necessary steps to obtain Eq. (\ref{nt}).

\begin{figure}
\includegraphics[width=\textwidth]{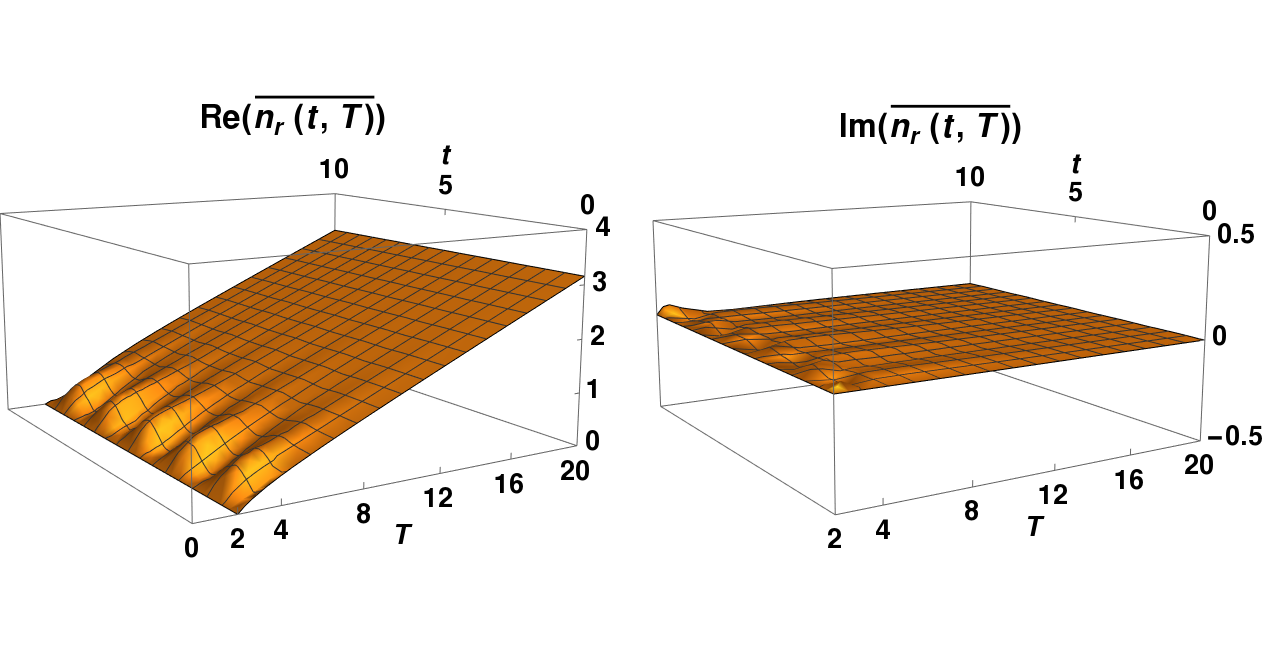}
\caption {Real and imaginary part of $\overline{n_(t,1/T)} $ as a function of time, $t$, and the temperature, $T=1/\beta$, for $|\Om|= 2 \pi$, $W=\pi$ and $V=\pi$.}\label{fig5}
\end{figure}

In Figure \ref{fig5}, the real and imaginary parts of $\overline{n_r(t,\beta)}$ of Eq. (\ref{nt}) are plotted as a function of the time, $t$, and the temperature, $T=1/\beta$. We have set $|\Om|= 2 \pi$, $W=\pi$ and $V=\pi$. The behaviour of the energy, $U_r(t,\beta)$ follows the same trend as $\overline{n_r(t,\beta)}$, scale by $\hbar |\Om|$. At low temperatures, it can observe a small imaginary contribution to the mean value of the energy, this can be interpreted as a sign that the system is unstable \cite{gadella1,gadella2}. At high temperatures, the effect of the perturbation is washed out.

\section{Conclusions}\label{conclusions}

In this work, we have investigated the dynamics of  pseudo-hermitian Hamiltonians at finite temperatures. With the aim of doing so,
we have summarised the essentials to compute the time evolution of these systems, i.e. density matrix and evolution operators, together with the extension of the Double-time temperature-dependent Green Function formalism to include this class of Hamiltonians. As an example, we have studied the Swanson model, both in the PT-symmetry and in the PT-broken symmetry phase.
We analyse the behaviour of the system, initially at equilibrium at a given temperature, when it is perturbed by a periodic interaction term. For the PT-symmetry phase, we have computed the solution in the linear approximation, and we have compared it with the exact solution. The comparison shows the advantages of the formalism. For the PT-broken symmetry phase, we compute the reduced density matrix on the states of $\Phi_+^\times$ to second order, and from it we discussed the thermodynamics properties of an inverted harmonic oscillator in presence of a periodic perturbation.

From the results presented in this work, we conclude that the Double Green Function formalism provides an adequate framework to study the evolution in time of non-hermitian systems at finite temperatures.

\section*{Acknowledgements}

This work was partially financed by CONICET, Argentina.

\section{Appendix A}\label{a0}

We shall prove that the operator 
\beqn 
\hat {\widetilde \rho}(t) = {\hat {\widetilde U}}(t,t_0)  \hat {\widetilde \rho}(t_0) {\hat {\overline U}}^\dagger(t,t_0),
\eeqn
with  $\hat {\widetilde \rho}(t_0)= \Upsilon^{-1}(t_0) \hat \rho_0 \Upsilon (t_0)$,
obeys the equation
\beqn 
{\bf i} \hbar \frac{\rm {d} \hat {\widetilde \rho}(t)}{\rm d t} = [\hat {\mathrm H}^\times, \hat {\widetilde \rho}(t)].
\eeqn 
To do so 

\beqn 
{\bf i} \hbar \frac{\rm {d} \hat {\widetilde \rho}(t)}{\rm d t} & = &
{\bf i} \hbar \frac{\rm {d} \hat {\widetilde U}(t,t_0)}{\rm d t} 
\hat {\widetilde \rho}(t_0) 
\hat {\overline U}^\dagger (t,t_0)+
{\bf i} \hbar 
\hat {\widetilde U}(t,t_0) \hat {\widetilde \rho}(t_0) 
\frac{\rm {d} \hat {\overline U}^\dagger(t_0)}{\rm d t} 
\nonumber \\
& = &
\hat {\mathrm H}^\times \hat {\widetilde U}(t,t_0)
\hat {\widetilde \rho}(t_0) 
\hat {\overline U}^\dagger (t,t_0)-
\hat {\widetilde U}(t,t_0) \hat {\widetilde \rho}(t_0)
\hat {\overline U}^\dagger (t,t_0)
\hat {\mathrm H}^\times
       \nonumber \\
& = & [\hat {\mathrm H}^\times, \hat {\widetilde \rho}(t)].
\eeqn 

Let us prove that
\beqn 
\langle \hat O \rangle = {\rm {Tr}}_{(\mathrm H^\times)} ( \hat {\widetilde \rho}(t) \hat O)={\rm {Tr}}_{(\mathfrak h)} ( \hat \rho(t) \hat o)= \langle \hat o \rangle.
\eeqn

\beqn 
\langle \hat O \rangle & = & {\rm {Tr}}_{(\mathrm H^\times)} ( \hat {\widetilde \rho}(t) \hat O) \nonumber \\
& = & {\rm {Tr}}_{(\mathrm H^\times)} ( {\widetilde U}(t,t_0) \hat {\widetilde \rho}_0(t) {\overline U}^\dagger (t,t_0) \hat O) \nonumber \\
& = &  {\rm {Tr}} ( {\widetilde U}(t,t_0) \Upsilon^{-1}(t_0) \hat {\rho}_0(t) \Upsilon(t_0) {\overline U}^\dagger (t,t_0) \hat O) \nonumber \\
& = &  {\rm {Tr}} ( {\widetilde U}(t,t_0) \Upsilon^{-1}(t_0) \hat {\rho}_0(t) \Upsilon(t_0) {\overline U}^\dagger (t,t_0) \Upsilon^{-1}(t)\hat o \Upsilon(t)) \nonumber \\
& = &  {\rm {Tr}} (\Upsilon^{-1}(t) {U}(t,t_0) \Upsilon(t_0)\Upsilon^{-1}(t_0) \hat {\rho}_0(t) \Upsilon(t_0) \Upsilon^{-1}(t_0){U}^\dagger (t,t_0) \Upsilon(t) \Upsilon^{-1}(t)\hat o \Upsilon(t)) \nonumber \\
& = &  {\rm {Tr}} ( {U}(t,t_0)  \hat {\rho}_0(t) {U}^\dagger (t,t_0) \hat o ) \nonumber \\
& = &  {\rm {Tr}} (  \hat {\rho}(t) \hat o ) \nonumber \\
\eeqn 

\section*{Appendix B}\label{a2}

We shall calculate the approximate solution of ${\overline{\hat x^2(t)}}$ and ${\overline{\hat p^2(t)}}$ from Eq. (\ref{ameanap}). To do so, we have to compute:

\beqn 
{\mathfrak h}_I(\tau) & = & \re^{{\bf i} {\mathfrak h_0} \tau } {\mathfrak h}_I \re^{{-\bf i} {\mathfrak h_0} \tau } \nonumber \\
          & = & \frac 12 V(t) \left( {\tilde a}^{\dagger 2} \re^{{\bf i} 2 \Omega \tau } +
          \aamt^2 \re^{-{\bf i} 2 \Omega \tau } + 2\aapt \aamt +1\right) \\
\hat x^2(t) & = & \re^{{\bf i} {\mathfrak h_0} t } \hat x^2 \re^{{-\bf i} {\mathfrak h_0} t }, \nonumber \\
          & = & \frac 12 \tilde b_0^2\left( {\tilde a}^{\dagger 2} \re^{{\bf i} 2 \Omega t } +
          \aamt^2 \re^{-{\bf i} 2 \Omega t} + 2\aapt \aamt +1\right),\\  
\hat p^2(t) & = & \re^{{\bf i} {\mathfrak h_0} t } \hat p^2 \re^{{-\bf i} {\mathfrak h_0} t }, \nonumber \\
          & = &- \frac 12 \frac{\hbar^2}{\tilde b_0^2}
          \left( {\tilde a}^{\dagger 2} \re^{{\bf i} 2 \Omega t } +
          \aamt^2 \re^{-{\bf i} 2 \Omega t } -( 2\aapt \aamt +1) \right).\\     
\eeqn 

Notice that

\beqn  
\langle [\aamt^2, {\mathfrak h}_I(\tau)] \rangle_0& = & 
{\rm{Tr}}(\re^{-\beta {\mathfrak h_0} } (2 \aapt \aamt+1) \re^{{\bf i} 2 \Omega t } )/Z,\nonumber \\
& = & \coth \left( \beta \frac {\hbar \Omega}{2} \right) \re^{{\bf i} 2 \Omega t } , \nonumber \\
\langle [{\tilde a}^{\dagger 2}, {\mathfrak h}_I(\tau)] \rangle_0& = & \coth \left( \beta \frac {\hbar \Omega}{2} \right) \re^{-{\bf i} 2 \Omega t }, \nonumber \\
\langle [\aapt \aamt, {\mathfrak h}_I(\tau)] \rangle_0& = & 0,
\eeqn
then 

\beqn  
\langle [ {\aamt}^2 \re^{-{\bf i} 2 \Omega t}+{\tilde a}^{\dagger 2} \re^{{\bf i} 2 \Omega t}, H_I(\tau) ] \rangle_0 =
-{\bf i} V(t) \coth (\beta \frac {\hbar \Omega}{2}) \sin(2 \Omega (t-\tau)). \nonumber \\
\eeqn

Finally:

\beqn 
{\rm {Tr}}(\hat x^2~\Delta \rho(t))=\frac 12  \tilde b_0^2  \coth \left( \beta \frac {\hbar \Omega}{2} \right) \int_{-\infty}^t ~V(\tau) \sin(2 \Omega (t-\tau)){\rm d} \tau, \nonumber \\
{\rm {Tr}}(\hat x^2~\Delta \rho(t))=- \frac 12 \frac{\hbar^2}{\tilde b_0^2} \coth \left (\beta \frac {\hbar \Omega}{2} \right) \int_{-\infty}^t ~V(\tau) \sin(2 \Omega (t-\tau)) {\rm d} \tau. \nonumber \\
\eeqn

\section*{Appendix C}\label{a1}

To compute $Z={\rm Tr} (\re^{-\beta \hbar \Omega} )$, we shall work with the faithful representation of the $su(1,1)$ algebra. The operators

\beqn 
K_+ = \frac {{a^\dagger}^2}2, ~~~
K_- = \frac {a^2}2, ~~~
K_0 = \frac 12 a^\dagger a + \frac 14,
\eeqn
obey the $su(1,1)$ algebra:

\beqn 
[K_\pm,K_0]= \pm K_\pm,~~~[K_-,K_+]= 2 K_0.
\eeqn 

The faithful representation is given by

\beqn  
K_+ \sim \left (  \begin{array} {cc}
                   0 & 1 \\ 0& 0
                  \end{array}\right ),~
K_- \sim \left (  \begin{array} {cc}
                   0 & 0 \\ -1 & 0
                  \end{array}\right ),~
K_+ \sim \left (  \begin{array} {cc}
                   \frac 12  & 0 \\ 0 & -\frac 12
                  \end{array}\right ).
\eeqn 
In general we want to express $\re^{A K_+ + B K_0 + C 2 K_0}$ as $\re^{\beta_+ K_+} \re^{\ln(\beta_0) 2 K_0 } \re^{\beta_- K_-}$, that is

\beqn  
\left (  \begin{array} {cc}
                  \cosh(r)+ \frac C r \sinh(r) & \frac A r \sinh(r) \\ -\frac B r \sinh(r) &  \cosh(r)- \frac C r \sinh(r)
\end{array}\right )
                  = 
\left (  \begin{array} {cc}
                   \beta_0- \frac{\beta_+ \beta_-}{\beta_0} & \frac {\beta_+} {\beta_0} \\ -\frac {\beta_-} {\beta_0} & \frac 1 {\beta_0}
\end{array}\right ),\nonumber \\
\eeqn 
where $r=\sqrt{C^2-A B}$.

Let us evaluate $U^{\dagger} \hat A U=U_I^{\dagger} \re^{{\bf i} H_0 t} \hat A \re^{-{\bf i} H_0 t} U_I$. We can proceed as before:

\beqn 
U_I(t) = \re^{-{\bf i} \left( \gamma K_+ + \gamma^* K_- + \gamma_0 2 K_0\right)}=
\re^{ \zeta_+ K_+}
\re^{ \ln(\zeta_0)~ 2 K_0}
\re^{ \zeta_-  K_-},
\eeqn 
with

\beqn 
\zeta_0 & = & \left ( \cos(d)+ {\bf i} \frac {\kappa_0} d (\sin(d) \right )^{-1}, \nonumber \\
\zeta_+ & = & -{\bf i} \frac {\kappa} d \sin(d) \zeta_0 , \nonumber \\
\zeta_- & = & -{\bf i} \frac {\kappa^*} d \sin(d) \zeta_0 , \nonumber \\
\eeqn  
where $d=\sqrt{\kappa_0^2-|\kappa|^2}$.

After some algebra

\beqn 
U^{\dagger} K_+ U & = & \re^{{\bf i} 2 \Omega t} \frac 1 {\zeta_0^2}  
\left( K_+ +\zeta_-^2 K_- - \zeta_- ~2 K_0\right) \nonumber \\
U^{\dagger} K_- U & = & \re^{-{\bf i} 2 \Omega t} \frac 1 {{\zeta_0^*}^2}  
\left( K_- +{\zeta_-^*}^2 K_+ -  \zeta_-^* ~ 2 K_0\right) \nonumber \\
U^{\dagger} 2 K_0 U & = & 
2\frac {\zeta_+} {{\zeta_0}^2}   K_+ + 2  \left( \frac {\zeta_+} {{\zeta_0}^2}\right)^* K_- +  \left( 1- 2 \frac {\zeta_- \zeta_+}{\zeta_0^2} \right)~ 2 K_0.
\eeqn


Finally, we calculate ${\rm Tr} (\rho_0 U^{-1} \hat A U)$ for $\hat A=K_+,~K_-$ and $2 K_0$:

\beqn 
{\rm Tr} (\rho_0 U^{\dagger} \hat A U)& = & {\rm Tr} (\rho_0 (c_+ K_++ c_- K_-+ c_0~ 2 K_0)) \nonumber \\
                                & = & {\rm Tr} (\rho_0 ~ c_0~ 2 K_0) \nonumber \\
                                & = & c_0 ~\frac 12 \coth \left(\frac{\beta  \Omega }{2}\right),
\eeqn
thus
\beqn 
{\rm Tr} (\rho_0 U^{\dagger} K_+ U ) & = & 
-\re^{{\bf i} 2 \Omega t} 
\frac {\zeta_-} {\zeta_0^2} ~ 
\frac 12 \coth \left(\frac{\beta  \Omega }{2}\right),
\nonumber \\
{\rm Tr} (\rho_0 U^{\dagger} K_- U ) & = & -\re^{-{\bf i} 2 \Omega t} \frac {\zeta_-^*} {{\zeta_0^*}^2} ~
\frac 12 \coth \left(\frac{\beta  \Omega }{2}\right), \nonumber \\
{\rm Tr} (\rho_0 U^{\dagger} 2 K_0 U ) & = & 
  \left( 1- 2 \frac {\zeta_- \zeta_+}{\zeta_0^2} \right)~ \frac 12 \coth \left(\frac{\beta  \Omega }{2}\right).
\eeqn

\section*{Appendix D} \label{a3}

Let us present the intermediate steps to obtain the expression for $\overline{n(t)}$. 

Following the work\cite{maru}, after the transformation to new variables, ${\mathfrak h}^\times$ can be written as

\beqn 
{\mathfrak{h}}^\times = - \frac 1 2 \hbar |\Om| \left(  \hat{u} \hat{v} +  \hat{v} \hat{u} \right),
\eeqn 
with
\beqn
{\hat u} = \sqrt{\frac{m |\Om|}{2 \hbar}  } \left( \hat{x}+ \frac{1}{m |\Om| }{\hat p} \right),
~~~~{\hat v} = \sqrt{\frac{m |\Om|}{2 \hbar}  } \left( \hat{x}-\frac{1}{m |\Om|} {\hat p} \right).
\label{uv2b}
\eeqn
In terms of the discrete eigenvalues and the corresponding generalised eigenvectors, the spectral representation can be written as 
\beqn
{\mathfrak{h}}^\times = {\mathfrak{h}}^+ +  {\mathfrak{h}}^-,
\eeqn 
with

\beqn
{\mathfrak{h}}^+ & = & ~ \sum_n ~ E_n |\phi_n^+ \rangle \langle \phi_n^- |, \nonumber \\
{\mathfrak{h}}^- & = & -\sum_n ~ E_n |\phi_n^- \rangle \langle \phi_n^+ |.
\eeqn 
At $t=t_0$:

\beqn
\rho_0 & = & \frac 1 Z ~\re^{-\beta~{\mathfrak{h}}^\times }, ~~~~
Z = {\rm{Tr}}_S (\rho_0).
\label{rho0}
\eeqn
As the energy of the resonant and anti-resonant states are complex conjugate pairs, $Z$ takes  values in $\mathbb{R}$\cite{pseudo1}. It is straightforward to verify that

\beqn
Z=\frac{1}{ {(2 ~\sin( |\Om| \beta/2))}^2}.
\eeqn
Considering that for $t>0$, the system's time evolution is given in terms of the resonant states, we shall work with a reduced matrix by making a trace over the anti-resonant states. The result is 

\beqn
\rho_r^{(0)} & = & \frac 1 Z ~\re^{-\beta~{\mathfrak{h}}^+ }, ~~~~Z_r = \frac{ 1}{2 \bf{i} ~\sin( |\Om| \beta/2)}.
\label{rho0r}
\eeqn
A similar result can be achieved by solving the equation for $\rho$ in the coordinate representation. That is 

\beqn 
-\frac{\partial~}{\partial \beta}\rho(x,x',\beta) =\left (- \frac{\hbar^2}{2 m}\frac{\partial^2 }{\partial x}-\frac{m |\Om|^2}{2} \hat  x^2 \right)\rho(x,x',\beta).
\eeqn 
The solution of the previous equation reads

\beqn 
\rho(x,x',\beta)=\sqrt{  \frac{m |\Om|}{2 \pi \hbar ~\sin(\beta |\Om|)}}~\re^{-\frac{m |\Om|}{2 \hbar ~\sin(\beta |\Om|)} \left( ( x^2+{x'}^2) \cos(\beta |\Om|)- 2~x x' \right)},\nonumber \\
\eeqn
and the associated partition function can be obtained as

\beqn 
Z_r=\int_{-\infty}^{\infty}~\rho(x,x,\beta)~{\rm{d x}}=  \frac{ 1}{2 {\bf{i}} ~\sin( |\Om| \beta/2)}.
\eeqn
The fact that $Z_r$ takes complex values is a consequence of working with an unstable system.

Though both the free energy and the entropy of the reduced system are complex, the internal energy takes real values:

\beqn
F_r & = & -T~\ln(Z_r)= -T \left( {\bf i} \frac \pi 2 - \ln \left( 2~\sin (|\Om|/(2 T)) \right )\right), \nonumber \\
S_r & = & -\frac{ \partial F_r}{\partial T}= {\bf i} \frac \pi 2 - \ln \left( 2~\sin (\beta |\Om|/2) \right )+ 
\frac 12 \hbar |\Om| {\rm {cotg}}\left ( \frac {\hbar |\Om|} {2 k_B T}\right), \nonumber \\
U_r & = & F_r + T S_r = \frac 12 \hbar |\Om| ~ {\rm {cotg}}\left ( \frac {\hbar |\Om|} {2 k_B T}\right), \nonumber \\
C_r & = & \frac{{\rm d} U_r}{{\rm d }T}=\left( \frac {\hbar |\Om|}{k_B T ~\sin(\hbar |\Om|/(2 T))} \right)^2, \nonumber \\
n_r & = & -T \frac{ \partial (\ln Z_r)}{\partial |\Om|} = \frac 12  ~ {\rm {cotg}} \left ( \frac {\hbar |\Om|} {2 k_B T}\right).
\eeqn 

\section*{Appendix E} \label{a3b}

Now, let us compute the correction to the matrix density when a perturbation interaction is switched on. To second order in the expansion

\beqn 
& & \rho(t) \approx \rho_0
- \frac {\bf i}{\hbar}~\re^{{\bf i} {\mathfrak h}_0^\times t/\hbar} ~\int_{-\infty}^t {\rm d} \tau  ~ [{\mathfrak h}_I^\times(\tau),\rho_0] 
~\re^{-{\bf i} {\mathfrak h}_0^\times t/\hbar}
\nonumber \\
& & ~~~+
\left( \frac {\bf i}{\hbar} \right)^2 \re^{{\bf i} {\mathfrak h}_0^\times t/\hbar} ~
\int_{-\infty}^t ~{\rm d} \tau~\int_{-\infty}^\tau ~{\rm d} \tau'
[{\mathfrak h}_I^\times(\tau),[{\mathfrak h}_I^\times(\tau'),\rho_0]] \re^{-{\bf i}~ {\mathfrak h}_0^\times t/\hbar}+...
\nonumber \\
\eeqn 
We shall take the trace over the states $\{ | \phi_n^- \rangle\}$. That is 

\beqn   
\rho_r(t) \approx \rho_r^{(0)}+  \rho_r^{(2)},
\eeqn 
with $\rho_r^{(0)}$ given in Eq. (\ref{rho0r}) and 

\beqn   
\rho_r^{(1)} = \sum_{n}~ \langle {\phi}_n^- | \rho_1 | {\phi}_n^- \rangle=0,
\eeqn
being

\beqn  
\rho_r^{(2)} & = & \sum_{n}~ \langle {\phi}_n^- | \rho_2 | {\phi}_n^- \rangle \nonumber \\
& = & 
-\re^{\uni {\mathfrak h}_+^\dagger t} \frac{\re^{-\beta {\mathfrak h}_+}}{Z} \re^{-\uni {\mathfrak h}_+ t}~{\mathfrak F}(t,\beta), ~~~Z= \frac{1}{( 2 \sin(\beta |\Om|/2))^2}.
\eeqn   

To obtain the mean value of the number of particles, we have to compute

\beqn   
\overline{n_r(t,\beta)} & \approx & \overline{n_r^{(0)}(\beta)}+ \overline{n_r^{(2)}(t,\beta)}.
\eeqn   

The second order correction to $\overline{n_r(t,\beta)} $ is given by

\beqn  
\overline{n_r^{(2)}(t,\beta)}& = & {\rm{Tr} }\left( -\frac 12 \{\hat u, \hat v\} \rho_r^{(2)} (t,\beta)\right)_S , 
\nonumber \\
& = &
\sum_n \langle {\phi}_n^- |
\frac 12 \{\hat u, \hat v\} \re^{\uni {\mathfrak h}_+^\dagger t} \frac{\re^{-\beta {\mathfrak h}_+}}{Z} \re^{-\uni {\mathfrak h}_+ t}
| {\phi}_n^+ \rangle
~{\mathfrak F}(t,\beta) \nonumber \\
& = & - \frac{1}{2 Z} S_0(\beta-2 \uni t) {\mathfrak F}(t,\beta).
\eeqn


\end{document}